\newif\ifconf
\conffalse
\conftrue 

\documentclass[11pt]{article}
\usepackage{amssymb}
\usepackage{amsthm}
\usepackage{amsmath}
\usepackage{fullpage}
\usepackage{epsfig}
\usepackage{verbatim}

\newcommand\remove[1]{}

\newcommand{\rnote}[1]{}
\newcommand{\jnote}[1]{}

\newcommand{\Lip}{\mathrm{Lip}}
\newcommand{\conv}{\mathrm{conv}}
\newcommand{\vol}{\mathrm{vol}}

\newcommand{\e}{\varepsilon}

\newcommand{\supp}{{{\mathrm{supp}}}}

\newcommand{\E}{\mathbb{E}}
\newcommand{\mytau}{2}
\newcommand{\roundup}[1]{{\lceil {#1} \rceil}}

\DeclareMathOperator{\diam}{diam}

\newtheorem{theorem}{Theorem}[section]
\newtheorem{lemma}[theorem]{Lemma}

\newtheorem{claim}[theorem]{Claim}

\newtheorem{definition}[theorem]{Definition}

\newtheorem{remark}{Remark}[section]

\newcommand{\myparagraph}[1]{\medskip\noindent{\bf #1.}}

\begin{document}

\date{}

\title{Measured descent: A new embedding method for finite
metrics}
\author{Robert Krauthgamer\thanks{
IBM Almaden Research Center, San Jose, CA 95120, USA.  Email: {\tt robi@almaden.ibm.com}}
\and James R. Lee\thanks{Computer Science Division, University of California,
Berkeley, CA~94720. Supported by NSF grant CCR-0121555 and an NSF
Graduate Research Fellowship. Email: {\tt jrl@cs.berkeley.edu}}
\and Manor Mendel\thanks{Seibel Center for Computer Science,
University of Illinois at Urbana-Champaign, Urbana, IL 61801, USA.
Email: {\tt mendelma@uiuc.edu} } \and Assaf Naor\thanks{ Microsoft
Research, One Microsoft Way, Redmond, WA 98052, USA. Email: {\tt
anaor@microsoft.com} } }

\maketitle

\begin{abstract}
We devise a new embedding technique, which we call {\em measured
descent}, based on decomposing a metric space locally, at varying
speeds, according to the density of some probability measure. This
provides a refined and unified framework for the two primary
methods of constructing Fr\'echet embeddings for finite metrics,
due to [Bourgain, 1985] and [Rao, 1999]. We prove that any
$n$-point metric space $(X,d)$ embeds in Hilbert space with
distortion $O(\sqrt{\alpha_X \cdot \log n})$, where $\alpha_X$ is
a geometric estimate on the decomposability of $X$. As an
immediate corollary, we obtain an  $O(\sqrt{(\log \lambda_X) \log
n})$ distortion embedding, where $\lambda_X$ is the doubling
constant of $X$. Since $\lambda_X\le n$, this result recovers
Bourgain's theorem, but when the metric $X$ is, in a sense,
``low-dimensional,'' improved bounds are achieved.

Our embeddings are volume-respecting for subsets of arbitrary
size. One consequence is the existence of $(k, O(\log n))$
volume-respecting embeddings for all $1 \leq k \leq n$, which is
the best possible, and answers positively a question posed by U.
Feige. Our techniques are also used to answer positively a
question of Y. Rabinovich, showing that any weighted $n$-point
planar graph embeds in $\ell_\infty^{O(\log n)}$ with $O(1)$
distortion. The $O(\log n)$ bound on the dimension is optimal, and
improves upon the previously known bound of $O((\log n)^2)$.
\end{abstract}

\section{Introduction}

 The theory of low-distortion embeddings of finite metric
spaces into normed spaces has attracted a lot of attention in
recent decades, due to its intrinsic geometric appeal, as well as
its applications in Computer Science. A major driving force in
this research area has been the quest for analogies between the
theory of finite metric spaces and the local theory of Banach
spaces. While being very successful, this point of view did not
always result in satisfactory metric analogues of basic theorems
from the theory of finite-dimensional normed spaces. An example of
this is Bourgain's embedding theorem~\cite{Bourgain85}, the
forefather of modern embedding theory, which states that every
$n$-point metric space embeds into a Euclidean space with
distortion $O(\log n)$. This upper bound on the distortion is
known to be optimal~\cite{LLR95}. Taking the point of view that
$\log n$ is a substitute for the dimension of an $n$-point metric
space (see~\cite{Bourgain85}; this approach is clearly natural
when applied to a net in the unit ball of some normed space), an
analogue of John's theorem~\cite{john} would assert that $n$-point
metrics embed into Hilbert space with distortion $O(\sqrt{\log
n})$. As this is not the case, the present work is devoted to a
more refined analysis of the Euclidean distortion of finite
metrics, and in particular to the role of a metric notion of
dimension.

We introduce a new embedding method, called {\em measured
descent}, which unifies and refines the known methods of
Bourgain~\cite{Bourgain85} and Rao~\cite{Rao99} for
constructing Fr\'echet-type embeddings (i.e. embeddings where
each coordinate is proportional to the distance from some subset of the
metric space).
Our method yields an embedding of any $n$-point metric space $X$
into $\ell_2$ with distortion
$O(\sqrt{\alpha_X \log n})$, where $\alpha_X$ is a geometric
estimate on the decomposability of $X$
(see Definition~\ref{def:modulus} for details).
As $\alpha_X \leq O(\log n)$, we obtain a
refinement of Bourgain's theorem, and when $\alpha_X$ is small
(which includes several important families of metrics)
improved distortion bounds are achieved.
This technique easily generalizes to produce embeddings which
preserve higher dimensional structures (i.e. not just distances
between pairs of points).  For instance, our embeddings can be
made volume-respecting in the sense of Feige (see
Section~\ref{sec:results}), and hence we obtain optimal
volume-respecting embeddings for arbitrary $n$-point spaces.

\myparagraph{Applications} In recent years, metric embedding has
become a frequently used algorithmic tool. For example, embeddings
into normed spaces have found applications to approximating the
sparsest cut of a graph \cite{LLR95,AR98, ARV04} and the bandwidth
of a graph \cite{Feige00,DV01}, and to distance
labeling schemes (see e.g. \cite[Sec.~2.2]{Ind01}).
The embeddings introduced in this paper refine our knowledge on these
problems, and in some cases improve the known algorithmic results.
For instance, they immediately imply an improved approximate max-flow/min-cut
theorem (and algorithm) for graphs excluding a fixed minor,
an improved algorithm for approximating the bandwidth of graphs whose metric
has a small doubling constant, and so forth.

\subsection{Notation} \label{sec:notation}

Let $(X,d)$ be an $n$-point metric space. We denote by $B(x,r) =
\{ y \in X : d(x,y) < r \}$ the open ball of radius $r$ about $x$.
For a subset $S \subseteq X$, we write $d(x,S) = \min_{y \in S}
d(x,y)$, and define $\diam(S) = \max_{x,y \in S} d(x,y)$. We recall
that the {\em doubling constant} of $X$, denoted $\lambda_X$, is
the least value $\lambda$ such that every ball in $X$ can be
covered by $\lambda$ balls of half the
radius~\cite{Lar67,Ass83,Luukk98,Heinonen01}. We say that a
measure $\mu$ on $X$ is {\em non-degenerate} if $\mu(x) > 0$ for
all $x \in X$. For a non-degenerate measure $\mu$ on $X$ define
$\Phi(\mu) = \max_{x \in X} \mu(X)/\mu(x)$ to be the {\em aspect
ratio} of $\mu$.

Let $(X,d_X)$ and $(Y,d_Y)$ be metric spaces. A mapping $f:X\to Y$
is called $C$-Lipschitz if $d_Y(f(x),f(y))\le C\cdot d_X(x,y)$ for
all $x,y\in X$.
The mapping $f$ is called $K$-bi-Lipschitz
if there exists a $C > 0$ such that
$$CK^{-1} \cdot d_X(x,y) \le d_Y(f(x),f(y)) \le C\cdot d_X(x,y),$$ for
all $x,y\in X$.
The least $K$ for which $f$ is $K$-bi-Lipschitz is called the
distortion of $f$, and is denoted $\mathrm{dist}(f)$. The least
distortion with which $X$ may be embedded in $Y$ is denoted
$c_Y(X)$. When $Y=L_p$ we use the notation
$c_Y(\cdot)=c_p(\cdot)$. Finally, the parameter $c_2(X)$ is called
the Euclidean distortion of $X$.

\myparagraph{Metric Decomposition} Let $(X,d)$ be a finite metric
space. Given a partition $P=\{C_1,\ldots,C_m\}$ of $X$,
we refer to the sets $C_i$ as {\em clusters}. We write ${\cal
P}_X$ for the set of all partitions of $X$. For $x\in X$ and a
partition $P\in {\cal P}_X$ we denote by $P(x)$ the unique cluster
of $P$ containing $x$. Finally, the set of all probability
distributions on ${\cal P}_X$ is denoted ${\cal D}_X$.

\begin{definition}[Padded decomposition]
\label{def:padded}
A (stochastic) decomposition of a finite metric space $(X,d)$ is a
distribution $\Pr \in \mathcal D_X$ over partitions of $X$.
Given $\Delta>0$ and $\e:X\to (0,1]$,
a $\Delta$-bounded $\varepsilon$-padded decomposition is one which
satisfies the following two conditions.
\begin{enumerate}
\item
For all $P \in \supp(\Pr)$, for all $C \in P$, $\diam(C)
\leq \Delta$. \item For all $x \in X$, $\Pr [B(x, \varepsilon(x)
\Delta) \nsubseteq P(x)] \leq \frac{1}{2}.$
\end{enumerate}
\end{definition}

We will actually need a collection of such decompositions, with
the diameter bound $\Delta>0$ ranging over all integral powers of
$2$ (of course the value 2 is arbitrary).

\begin{definition}[Decomposition bundle]
\label{def:bundle} Given a function $\e:X\times {\mathbb Z}\to
(0,1]$, an $\varepsilon$-padded decomposition bundle on $X$ is a
function $\beta : \mathbb Z \to \mathcal D_X$, where for every $u
\in \mathbb Z$, $\beta(u)$ is a $2^u$-bounded
$\varepsilon(\cdot,u)$-padded stochastic decomposition of $X$.
\end{definition}

Finally, we associate to every finite metric space an important
``decomposability'' parameter $\alpha_X$. (See \cite{cluster} for
relationships to other notions of decomposability.)

\begin{definition}[Modulus of padded decomposability] \label{def:modulus}
The modulus of padded decomposability of a finite metric space $(X,d)$ is
defined as
$$
\alpha_X = \inf \{ \alpha : \textrm{there exists an
$\varepsilon$-padded decomposition bundle on $X$ with
$\varepsilon(x,u)\equiv 1/\alpha$} \}.$$
\end{definition}

It is known that $\alpha_X=O(\log n)$
\cite{LinialSaks,Bartal96}, and furthermore $\alpha_X=O(\log
\lambda_X)$ \cite{GKL03}.
Additionally, if $X$ is the shortest-path metric on an $n$-point
constant-degree expander, then $\alpha_X = \Omega(\log n)$
\cite{Bartal96}. For every metric space $X$ induced by an
edge-weighted graph which excludes $K_{r,r}$ as a minor, it is
shown in the sequence of papers \cite{KPR93, Rao99, FT03} that
$\alpha_X = O(r^2)$.

\myparagraph{Volume-respecting embeddings}
We recall the notion of {\em volume-respecting embeddings}, which
was introduced by Feige~\cite{Feige00} as a tool in his study of
the graph bandwidth problem. Let $S \subseteq X$ be a $k$-point
subset of $X$. We define its volume by
$$
\vol(S)=\sup\{\vol_{k-1}(\conv(f(S))):\ f:S\to L_2 \textrm{ is
1-Lipschitz}\},
$$
where for $A\subseteq L_2$, $\conv(A)$ denotes its convex hull and
the $(k-1)$-dimensional volume above is computed with respect to
the Euclidean structure induced by $L_2$. A mapping $f:X\to L_2$
is called $(k,\eta)$-volume-respecting if it is $1$-Lipschitz and
for every $k$-point subset $S\subset X$,
$$
\left[\frac{\vol(S)}{\vol_{k-1}(\conv(f(S)))}\right]^{\frac{1}{k-1}}\le\eta.
$$
It is easy to see that a 1-Lipschitz map $f : X \to L_2$ has
distortion $D$ if and only if it is $(2, D)$-volume-respecting.
Thus the volume-respecting property is a generalization of
distortion to larger subsets.

\subsection{Results}
\label{sec:results}
The following theorem refines Bourgain's result in terms of
 the decomposability parameter.

\begin{theorem}[Padded embedding theorem]
\label{thm:pad} For every $n$-point metric space $(X,d)$, and
every $1\le p\le \infty$,
\begin{equation} \label{eq:c_p}
c_p(X)\le O(\alpha_X^{1-1/p} (\log n)^{1/p}).
\end{equation}
\end{theorem}

The proof appears in Sections~\ref{sec:outline}
and~\ref{section:bourgain}. Since $\alpha_X= O(\log
\lambda_X)$, it implies in particular that $c_2(X)\le
O\bigl(\sqrt{(\log \lambda_X)\cdot \log |X|}\bigr)$, for any
metric space $X$. This refines Bourgain's embedding
theorem~\cite{Bourgain85}, and improves upon previous embeddings
of doubling metrics~\cite{GKL03}. It is tight for
$\lambda_X=O(1)$~\cite{Laakso,Lang,GKL03}, and
$\lambda_X=n^{\Omega(1)}$~\cite{LLR95}. The question of whether
this bound is tight up to a constant factor for the range
$\lambda_X\in\{c_1,\ldots, |X|^{c_2}\}$, where $c_1\in
\mathbb{N}$, $0<c_2<1$ are some constants, is an interesting open
problem.

For $1 \leq p < 2$, the bound $O(\sqrt{\alpha_X \log n})$ is
better than \eqref{eq:c_p}, and thus, in these cases, it makes
sense to construct the embedding first into $L_2$.

\medskip

A more careful analysis of the proof of Theorem~\ref{thm:pad}
yields the following result, proved in
Section~\ref{section:volume}, which answers a question posed by
Feige in~\cite{Feige00}.

\begin{theorem}[Optimal volume-respecting
embeddings]\label{thm:feige} Every $n$-point metric space $X$
admits an embedding into $L_2$ which is $(k, O(\sqrt{\alpha_X \log
n}))$ volume-respecting for every $2\le k\le n$.
\end{theorem}

Since $\alpha_X = O(\log n)$, this provides $(k, O(\log
n))$-volume-respecting embeddings for every $2 \leq k \leq n$.
This is optimal; a matching lower bound is given in~\cite{KLM} for
all $k<n^{1/3}$.  We note that the previous best bounds were due
to Feige \cite{Feige00}, who showed that a variant of Bourgain's
embedding achieves distortion $O(\sqrt{\log n}\cdot \sqrt{\log n +
k \log k})$ (note that this is $\Omega(\sqrt{n})$ for large values
of $k$), and to Rao who showed that $O((\log n)^{3/2})$ volume
distortion is achievable for all $1 \leq k \leq n$ (this follow
indirectly from \cite{Rao99}, and was first observed in
\cite{Gupta-thesis}).

This also improves the dependence on $r$ in Rao's
volume-respecting embeddings of $K_{r,r}$-excluded metrics, from
$(k,O(r^2 \sqrt{\log n}))$, due to \cite{Rao99,FT03}, to $(k,O(r
\sqrt{\log n}))$.\footnote{This bound is tight for the path even
for $k=3$, see \cite{DV01,KLM}.} As a corollary, we obtain an
improved $O(r\sqrt{\log n})$- approximate max-flow/min-cut
algorithm for graphs which exclude $K_{r,r}$ as a minor.

\myparagraph{$\ell_\infty$ embeddings} It is not difficult to see
that every $n$-point metric space $(X,d)$ embeds isometrically
into $\ell_\infty^n$ via the map $y \mapsto \left\{ d(x,y)
\right\}_{x\in X}$. And for some spaces, like the shortest-path
metrics on expanders, or on the $\log n$-dimensional hypercube
(see, e.g. \cite{LMN04}), it is known that $n^{\Omega(1)}$
dimensions are required to obtain any map with $O(1)$ distortion.
On the other hand, a simple variant of Rao's embedding shows that
every planar metric $O(1)$-embeds into $\ell_\infty^{O((\log
n)^2)}$. Thus the dimension required to embed a family of metrics
into $\ell_\infty$ with low distortion is a certain measure of the
family's complexity (see \cite{Mat01}).

In
Section~\ref{section:planar} we use a refinement of measured
descent to prove the following theorem, which answers positively a
question posed by Y. Rabinovich~\cite{jiriproblems}, and
improves Rao's result to obtain the optimal bound.

\begin{theorem}\label{thm:planar} Let $X$ be an $n$-point edge-weighted planar
graph, equipped with the shortest path metric. Then $X$ embeds
into $\ell_\infty^{O(\log n)}$ with $O(1)$ distortion.
\end{theorem}

The $O(\log n)$ bound on the dimension is clearly optimal (by
simple volume arguments). Furthermore, this result is stronger
than the $O(\sqrt{\log n})$ distortion bound on Euclidean
embeddings of planar metrics, due to Rao~\cite{Rao99}. The
embedding is produced by ``derandomizing'' both the decomposition
bundle of \cite{Rao99, KPR93} and the proof of measured descent
(applied to this special decomposition bundle).

\subsection{The main technical lemmas} \label{sec:outline}

The following lemma is based on a decomposition of~\cite{CKR01},
with the improved analysis of~\cite{FHRT03, FRT03}.  The extension
to general measures was observed in \cite{LN03-unified}. Since
this lemma is central to our techniques, its proof is presented in
Section~\ref{section:bourgain} for completeness.
Throughout, $\log x$ denotes the natural logarithm of $x$.
\begin{lemma}
\label{lem:FRT} Let $(X,d)$ be a finite metric space and let $\mu$
be any non-degenerate measure on $X$. Then there exists an
$\varepsilon(x, u)$-padded decomposition bundle on $X$ where
\begin{eqnarray}\label{eq:defepsilon}
\varepsilon(x,u) = \left[16+16 \log
\frac{\mu(B(x,2^u))}{\mu(B(x,2^{u-3}))}\right]^{-1}.\end{eqnarray}
\end{lemma}

\begin{remark}
In~\cite{volberg} it was shown that $X$ admits a doubling measure,
i.e. a non-degenerate measure $\mu$ such that for every $x\in X$
and every $r>0$ we have
$\frac{\mu(B(x,2r))}{\mu(B(x,r))}=\lambda_X^{O(1)}$. We thus recover
the fact, first proved in~\cite{GKL03}, that for every metric
space $X$, $\alpha_X=O(\log \lambda_X)$. In particular, for every
$d$-dimensional normed space $Y$, $\alpha_Y = O(d)$. In
\cite{CCGGP}, it is argued that $\alpha_Y = \Omega(d)$ when $Y =
\ell_1^d$.  The same lower bound was shown to hold for every $d$
dimensional normed space $Y$ in~\cite{cluster}.

\end{remark}

\remove{Using a simple reduction (see \cite{unified}), it follows
that every two-dimensional Riemannian manifold $M$ of genus $g$
has $\alpha_M = O(g)$.  Since every $n$-point metric space embeds
isometrically into a two-dimensional Riemannian manifold of genus
$O(n^3)$, there are surfaces $M$ of genus $g$ for which we also
have $\alpha_M = \Omega(\log g)$.}

\myparagraph{The main embedding lemma} Let $(X,d)$ be a finite
metric space, and for $\e:X\times \mathbb Z\to \mathbb R$ define
for all $x,y\in X$,
$$
\delta_\e(x,y)=\min \left\{ \e(x,u):\ u\in {\mathbb Z} \mbox{
\;and \;} \frac{d(x,y)}{32} \le 2^u \le \frac{d(x,y)}{2} \right\}.
$$

Given a
non-degenerate measure $\mu$ on $X$ denote for $x,y\in X$:
\begin{equation}\label{eq:def:V}
V_\mu(x,y)=\max\left\{ \log\frac{\mu (B(x,2d(x,y)))}{\mu
(B(x,d(x,y)/512))},\log\frac{\mu (B(y,2d(x,y)))}{\mu
(B(y,d(x,y)/512))}\right\}.
\end{equation}

In what follows we use the standard notation $c_{00}$ for the
space of all finite sequences of real numbers.
The following result is the main embedding lemma of this paper.

\begin{lemma}[Main embedding lemma]
\label{lem:main} Let $X$ be an $n$-point metric space, $\mu$ a
non-degenerate measure on $X$, and $\beta : \mathbb Z \to \mathcal
D_X$ an $\varepsilon(x,u)$-padded decomposition bundle on $X$.
Then there exists a map $\varphi : X \to c_{00}$ such that for
every $1\le p \leq \infty$ and for all distinct $x,y \in X$,
$$
\left[V_\mu(x,y)\right]^{1/p}\cdot \min\{\delta_{\varepsilon}(x,
y),\delta_{\varepsilon}(y, x)\}
 \leq \frac{||\varphi(x) - \varphi(y)||_p}{d(x,y)} \leq C \left[\log
 \Phi(\mu)\right]^{1/p}.
$$
Here $C$ is a universal constant.
\end{lemma}

Using Lemma~\ref{lem:main} we are in a position to prove
Theorem~\ref{thm:pad}. We start with the following simple
observation, which bounds $V_\mu(x,y)$ from below, and which will
be used several times in what follows.

\begin{lemma}\label{lem:golden} Let $\mu$ be any non-degenerate
measure on $X$ and $x,y\in X$, $x\neq y$. Then
$$
\max\left\{ \frac{\mu (B(x,2d(x,y)))}{\mu
(B(x,d(x,y)/2))},\frac{\mu (B(y,2d(x,y)))}{\mu
(B(y,d(x,y)/2))}\right\}\ge 2.
$$
\end{lemma}

\begin{proof}
Assume without loss that $\mu (B(x,2d(x,y))) \le \mu
(B(y,2d(x,y)))$. Noticing that the two balls $B(x,d(x,y)/2)$ and
$B(y,d(x,y)/2)$ are disjoint, and that both are contained in
$B(x,2d(x,y))$; the proof follows.
\end{proof}


\begin{proof}[Proof of Theorem~\ref{thm:pad}]
Fix $p \in [1,\infty]$ and let $\mu=|\cdot|$ be the counting
measure on $X$. Let $\e(x,u)$ be as in \eqref{eq:defepsilon}, and
observe that in this case for all $x,y\in X$ we have
$\delta_\e(x,y)\ge [16+16V_\mu(x,y)]^{-1}$. Applying
Lemma~\ref{lem:main} to the decomposition bundle of
Lemma~\ref{lem:FRT} we get a mapping $\varphi_1:X\to L_p$ such
that for all $x,y\in X$,
$$
\frac{\left[V_\mu(x,y)\right]^{1/p}}{16+16V_\mu(x,y)}
 \leq \frac{||\varphi_1(x) - \varphi_1(y)||_p}{d(x,y)} \leq C(\log n)^{1/p}.
$$
On the other hand, Lemma~\ref{lem:main} applied to the
decomposition bundle ensured by the definition of $\alpha_X$
yields a mapping $\varphi_2:X\to L_p$ for which
$$
\frac{\left[V_\mu(x,y)\right]^{1/p}}{\alpha_X}
 \leq \frac{||\varphi_2(x) - \varphi_2(y)||_p}{d(x,y)} \leq C(\log n)^{1/p}.
$$

Finally, for $\varphi = \varphi_1 \oplus \varphi_2$ we have
\begin{eqnarray*}
\frac{||\varphi(x) - \varphi(y)||_p^p}{d(x,y)^p} \geq
\frac{V_\mu(x,y)}{[16+16V_\mu(x,y)]^p}+\frac{V_\mu(x,y)}{\alpha_X^p}\ge
\Omega\left(\frac{1}{\alpha_X^{p-1}}\right),
\end{eqnarray*}
where we have used the fact that Lemma~\ref{lem:golden} implies
that $V_\mu(x,y)\ge \Omega(1)$.
\end{proof}

\section{Measured descent}
\label{section:bourgain}

In this section, we prove the main embedding lemma and exhibit the
existence of optimal volume-respecting embeddings. We use
decomposition bundles to construct random subsets of $X$, the
distances from which are used as coordinates of an embedding into
$c_{00}$. As the diameters of the decompositions become smaller,
our embedding ``zooms in'' on the increasingly finer structure of
the space. Our approach is heavily based on the existence of good
decomposition bundles; we thus start by proving
Lemma~\ref{lem:FRT}, which is essentially contained
in~\cite{FRT03}.

\begin{proof}[Proof of Lemma~\ref{lem:FRT}] By approximating
the values $\{\mu(x)\}_{x\in X}$ by rational numbers and
duplicating points, it is straightforward to verify that it is
enough to prove the required result for the counting measure on
$X$, i.e. when $\mu(S) = |S|$.

Let $\Delta = 2^u$ for some $u \in \mathbb Z$. We now describe the
distribution $\beta(u)$. Choose, uniformly at random, a
permutation $\pi$ of $X$ and a value $\alpha \in
[\frac{1}{4},\frac{1}{2}]$. For every point $y \in X$, define a cluster
$$
C_y = B(y,\alpha\Delta)\setminus \bigcup_{z:\pi(z)<\pi(y)}
B(z,\alpha\Delta).$$ In words, a point $x\in X$ is assigned to
$C_y$ where $y$ is the minimal point according to $\pi$ that is
within distance $\alpha\Delta$ from $x$.

Clearly the set $P = \{C_y\}_{y \in X}$
constitutes a partition of $X$.  Furthermore, $C_y \subseteq B(y,
\alpha \Delta)$, thus $\diam(C_y) \leq \Delta$, so requirement (1)
in Definition \ref{def:padded} is satisfied for every partition
$P$ arising from this process. It remains to prove requirement (2).

Fix a point $x \in X$ and some value $t \le \Delta/8$. Let $a =
|B(x, \Delta/8)|$, $b = |B(x, \Delta)|$, and arrange the points
$w_1, \ldots, w_{b} \in B(x, \Delta)$ in increasing distance from
$x$. Let $I_k = [d(x, w_k)-t, d(x,w_k)+t]$ and write $\mathcal
E_k$ for the event that $\alpha\Delta \le d(x,w_k)+t$ and $w_k$ is
the minimal element according to $\pi$ such that $\alpha\Delta \ge
d(x,w_k)-t$. Note that if $w_k \in B(x, \Delta/8)$, then
$\Pr[\mathcal E_k] = 0$ since in this case
$d(x,w_k)+t<\Delta/8+t\le \Delta/4\le \alpha\Delta$. We claim that
the event $\{ B(x,t)\nsubseteq P(x)\}$ is contained in the event
$\bigcup_{k=1}^b \mathcal E_k$. Indeed, let $w_j$ be the minimal
element according to $\pi$ such that $C_{w_j}\cap B(x,t)\neq
\emptyset$. It follows that $d(w_j,x)<\alpha\Delta+t$.
Furthermore, $d(x,w_j)\ge \alpha\Delta-t$, since otherwise $
B(w_j,\alpha\Delta)\supseteq B(x,t)$, implying that $P(x)=C_{w_j}
\supseteq B(x,t)$. Hence there exists $k$ for which the event
$\mathcal E_k$ occurs.

Now
\begin{eqnarray*}
\Pr[B(x,t) \nsubseteq P(x)] \le  \sum_{k=a+1}^{b} \Pr[\mathcal
E_k] &=& \sum_{k=a+1}^b \Pr[\alpha \Delta \in I_k] \cdot
\Pr[\mathcal E_k\,|\,\alpha \Delta \in I_k] \\ &\leq&
\sum_{k=a+1}^b \frac{2t}{\Delta/4} \cdot \frac{1}{k} \leq
\frac{8t}{\Delta} \left(1+\log \frac{b}{a}\right),
\end{eqnarray*}
where we used the fact that $\Pr[\mathcal E_k\,|\,\alpha \Delta
\in I_k]\le \Pr[j<k\implies \pi(j)>\pi(k)]=1/k$. Setting $t =
\varepsilon(x, u) \Delta\le \Delta/8$, where $\e(x,u)$ is as in
\eqref{eq:defepsilon}, the righthand side is at most
$\frac{1}{2}$, proving requirement (2) in Definition
\ref{def:padded}.
\end{proof}

\subsection{Proof of main embedding lemma}\label{section:proofbourgain}

We first introduce some notation.
Define two intervals of integers
$I,T\subseteq \mathbb Z$ by
$$
I=\{-6,-5,-4,-3,-2,-1,0,1,2,3\}\quad \mathrm{and}\quad
T=\{0,1,\ldots,\lceil\log_\mytau \Phi(\mu)\rceil\}.
$$

 For $t> 0$ write $ \kappa(x,t) = \max \{ \kappa \in \mathbb Z:
\mu(B(x,2^\kappa)) < \mytau^t\}$. For each $u\in \mathbb Z$ let
$P_u$ be chosen according to the distribution $\beta(u)$.
Additionally, for $u\in \mathbb Z$ let $\{\sigma_u(C):\ C\subseteq
X\}$ be i.i.d. symmetric $\{0,1\}$-valued Bernoulli random
variables. We assume throughout the ensuing argument that the
random variables $\{\sigma_u(C):\ C\subseteq X,\ u\in \mathbb
Z\}$, $\{P_u:\ u\in \mathbb Z\}$ are mutually independent. For
every $t\in T$ and $i\in I$ define a random subset $W^i_t\subseteq
X$ by
$$
W^i_t=\{x\in X:\
\sigma_{{\kappa(x,t)-i}}(P_{{\kappa(x,t)-i}}(x))=0\}.
$$

Our random embedding $f:X\to c_{00}$ is defined by
$f(x)=(d(x,W^i_t):\ i\in I,\ t\in T)$.
In the sequel, we assume that $p < \infty$;
the case $p=\infty$ follows similarly.
Since each of the
coordinates of $f$ is Lipschitz with constant $1$, we have for all
$x,y\in X$,
\begin{eqnarray}\label{eq:upper}
\|f(x)-f(y)\|_p^p\le |I|\cdot|T|\cdot d(x,y)^p\le 50[\log
\Phi(\mu)]\, d(x,y)^p.
\end{eqnarray}
The proof will be complete once we show that for all $x,y\in X$
\begin{eqnarray}\label{eq:lower}
\E \|f(x)-f(y)\|_p^p \ge [\Omega(\min\{\delta_{\varepsilon}(x,y),\delta_\varepsilon(y,x)\} \cdot
d(x,y))]^p \cdot V_\mu(x,y).
\end{eqnarray}
Indeed, denote by $(\Omega,\Pr)$ the probability space on which the
above random variables are defined, and consider the space
$L_p(\Omega,c_{00})$, i.e. the space of all $c_{00}$ valued random
variables $\zeta$ on $\Omega$ equipped with the $L_p$ norm
$\|\zeta\|_p=(\E\|\zeta\|_p^p)^{1/p}$. Equations \eqref{eq:upper}
and \eqref{eq:lower} show that the mapping $x\mapsto f(x)$ is the
required embedding of $X$ into $L_p(\Omega,c_{00})$. Observe that all
the distributions are actually finitely supported, since $X$ is
finite, so that this can still be viewed as an embedding into
$c_{00}$. See Remark~\ref{remark:chernoff} below for more
details.

\medskip

To prove \eqref{eq:lower} fix $x,y\in X$, $x\neq y$. Without loss of
generality we may assume that the maximum in \eqref{eq:def:V} is attained by
the first term, namely,
$ \frac{\mu (B(x,2d(x,y)))}{\mu(B(x,d(x,y)/512))}
  \ge \frac{\mu (B(y,2d(x,y)))}{\mu(B(y,d(x,y)/512))}$.
Using Lemma~\ref{lem:golden}, it immediately follows that
\begin{equation}\label{eq:noloss}
  \frac{\mu (B(x,2d(x,y)))}{\mu(B(x,d(x,y)/512))}
  \ge 2.
\end{equation}

Setting $R=\frac14 d(x,y)$, denote
for $i\in \mathbb Z$, $s_i= \log_\mytau \mu(B(x,2^iR))$.
We next extend some immediate bounds on $\kappa(x,t)$ (in terms of $R$)
to any nearby point $z\in B(x,R/256)$.

\begin{claim}\label{claim:controlkappa}
For $i\in I$ and all $t\in \mathbb Z\cap[s_{i-1}, s_i]$, every
$z\in B(x,R/256)$ satisfies $ \frac{R}{8}\le 2^{\kappa(z,t)-i}<
\frac{5R}{4}. $
\end{claim}
\begin{proof}
By definition, $\mu\left(B(z, 2^{\kappa(z,t)})\right) < \mytau^t
\leq \mu\left(B(z, 2^{\kappa(z,t)+1})\right).$ For the upper
bound, we have
$$
\mu\left(B(x, 2^{\kappa(z,t)} - R/256)\right) \leq \mu\left(B(z,
2^{\kappa(z,t)})\right)  < \mytau^t \leq \mytau^{s_i}=
\mu\left(B(x,2^iR)\right),$$ implying that $2^{\kappa(z,t)} -
\frac{R}{256} < 2^iR$, which yields $2^{\kappa(z,t)-i}<
\frac{5R}{4}$. For the lower bound, we have
$$
\mu\left(B(x, 2^{\kappa(z,t) + 1} + R/256)\right) \geq
\mu\left(B(z, 2^{\kappa(z,t)+1})\right) \geq \mytau^t \geq
\mytau^{s_{i-1}}= \mu\left(B(x, 2^{i-1}R)\right).$$ We conclude that
$2^{\kappa(z,t)+1} + \frac{R}{256}\ge 2^{i-1}R$, which implies
that $\frac{R}{8}\le 2^{\kappa(z,t)-i}$.
\end{proof}

Consider the following  events
\begin{quote}
\begin{enumerate}
\item $\mathcal E_1=\left\{d(x, X \setminus P_{u}(x)) \geq
\delta_{\varepsilon}(x,y) \frac{R}{8}\ \mathrm{for\  all}\  u
\in \mathbb Z\ \mathrm{with}\  2^u \in [R/8, 5R/4]\right\}$,
\item
$\mathcal E_2=\left\{ \sigma_u(P_{u}(x)) = 1\ \mathrm{for\ all}\
u \in \mathbb Z\  \mathrm{with}\  2^u \in [R/8, 5R/4]\right\}$,
\item $\mathcal E_3= \left\{ \sigma_u(P_{u}(x)) = 0\ \mathrm{for\
all}\  u \in \mathbb Z\  \mathrm{with}\  2^u \in [R/8,
5R/4]\right\}$,
\item $\mathcal
E_{i,t}^{\mathrm{big}}=\left\{d(y,W^i_t)\ge \frac{1}{512}
\delta_{\varepsilon}(x,y) R\right\}$,
\item $\mathcal
E_{i,t}^{\mathrm{small}}= \left\{d(y,W^i_t)< \frac{1}{512}
\delta_{\varepsilon}(x,y) R\right\}$.
\end{enumerate}
\end{quote}

The basic properties of these events are described in the
following claim.

\begin{claim}\label{claim:probabilistic} The following assertions
hold true:
\begin{enumerate}
\setlength{\itemsep}{0in}
\renewcommand{\theenumi}{\rm{(\alph{enumi})}}
\item\label{item:bigprob} $\Pr[\mathcal E_1], \Pr[\mathcal E_2],
\Pr[\mathcal E_3] \geq 2^{-4}$.
\item\label{item:firstind}
For all $i \in I$ and $t\in \mathbb Z\cap [s_{i-1},s_i]$, the event
$\mathcal E_3$ is independent of $\mathcal E_{i,t}^{\mathrm{big}}$.
\item\label{item:secondind}
For all $i \in I$ and $t\in \mathbb Z\cap [s_{i-1},s_i]$, the event
$\mathcal E_2$ is independent of $\mathcal E_1\cap \mathcal E_{i,t}^{\mathrm{small}}$.
\item\label{item:inset} For all $i \in I$ and $t\in \mathbb Z\cap [s_{i-1},s_i]$,
if the event $\mathcal E_3$ occurs then $x\in W_i^t$.
\item\label{item:padded} For all $i \in I$ and $t\in\mathbb Z\cap [s_{i-1},s_i]$,
if the event $\mathcal E_1\cap \mathcal E_2$ occurs then
$d(x, W_i^t)\ge \frac{1}{256} \delta_{\varepsilon}(x,y) R$.
\end{enumerate}
\end{claim}
\begin{proof}
For the first assertion, fix $u$ such that $2^u \in [R/8, 5R/4]$.
Since $\delta_{\varepsilon}(x,y) \leq \varepsilon(x,u)$,
and $P_u$ is chosen from $\beta(u)$ which is $\varepsilon(x,u)$-padded,
$\Pr[d(x, X\setminus P_{u}(x)) \geq \delta_\varepsilon(x,y) 2^u] \geq \frac12$.
In addition, $\Pr[\sigma_u(P_u(x)) = 1] =\Pr[\sigma_u(P_u(x)) =
0]= 1/2$. Furthermore, the number of relevant values of $u$ in
each of the events $\mathcal E_1$, $\mathcal E_2$, $\mathcal E_3$ is at
most four for each event, and the outcomes for different values of
$u$ are mutually independent. This implies assertion \ref{item:bigprob}.

%
To prove the second and third assertions note
that for $2^u \in [R/8, 5R/4]$, we always have $\diam(P_u(x)) \leq
\frac{5R}{4} <\frac12\,d(x,y)$, and thus $P_u(x) \neq P_u(y)$.
Furthermore, for every $z\in B(y,\frac{1}{512} \delta_{\varepsilon}(x,y) R)$,
we always have $d(x,z) \ge 3R > \diam(P_u(x))+\diam(p_u(z))$,
thus $P_u(x) \neq P_u(z)$, and the choices of $\sigma_u(P_u(x))$
and $\sigma_u(P_u(z))$ are independent.
It follows that the value of $\sigma_u(P_u(x))$ is independent of
the data determining whether
$d(y,W^i_t) < \frac{1}{512} \delta_{\varepsilon}(x,y) R$,
which proves \ref{item:firstind}.
Assertion \ref{item:secondind} follows similarly observing that
$\sigma_u(P_u(x))$ is independent also of the value of $d(x,X\setminus P_u(x))$.

To prove the last two assertions fix $i \in I$ and $t\in\mathbb Z\cap [s_{i-1},
s_i]$. An application of Claim~\ref{claim:controlkappa} to
$z=x$ shows that $2^{\kappa(x,t)-i}\in [R/8,5R/4]$. Now, by the
construction of $W^i_t$, if $\mathcal E_3$ occurs then $x\in W^i_t$;
this proves assertion \ref{item:inset}. Finally,
fix any $z\in B\left(x, \frac{1}{256}\delta_{\varepsilon}(x,y) R\right)$.
Since $z \in B(x,R/256)$,
Claim~\ref{claim:controlkappa} implies that $2^{\kappa(z,t)-i} \in
[R/8, 5R/4]$. The event $\mathcal E_1$ implies that for $u=\kappa(z,t)-i$,
$P_u(x) = P_u(z)$, and thus
$\sigma_{\kappa(z,t)-i}(P_{\kappa(z,t)-i}(z)) =
\sigma_{\kappa(z,t)-i}(P_{\kappa(z,t)-i}(x))$.
Now $\mathcal E_2$ implies that the latter quantity is $1$,
and hence $z \notin W^i_t$. Assertion \ref{item:padded} follows.
\end{proof}

We can now conclude the proof of Lemma~\ref{lem:main}.
Fix $i \in I$ and $t\in[s_{i-1},s_i]$. By assertions \ref{item:inset} and
\ref{item:padded},
if either of the (disjoint) events $\mathcal E_3 \cap
\mathcal{E}_{i,t}^{\mathrm{big}}$ and $\mathcal E_1\cap \mathcal
E_2\cap  \mathcal{E}_{i,t}^{\mathrm{small}}$ occurs then
$|d(x,W^i_t)-d(y,W^i_t)|\ge \frac{1}{512}
\delta_{\varepsilon}(x,y) R$.
The probability of this is
$ \Pr[\mathcal E_3]\cdot\Pr[\mathcal{E}_{i,t}^{\mathrm{big}}]+\Pr[\mathcal
E_2]\cdot\Pr[\mathcal E_1\cap \mathcal{E}_{i,t}^{\mathrm{small}}]
\ge 2^{-4} \Pr[\mathcal E_1] =\Omega(1)$,
where we have used assertions \ref{item:bigprob}, \ref{item:firstind}
and \ref{item:secondind}, and the fact that
$ \mathcal{E}_{i,t}^{\mathrm{big}} \cup (\mathcal E_1\cap \mathcal{E}_{i,t}^{\mathrm{small}}) \supseteq \mathcal E_1$.
It follows that
$\E|d(x,W^i_t)-d(y,W^i_t)|^p=[\Omega(\delta_\varepsilon(x,y)\cdot d(x,y))]^p$,
and hence
\begin{eqnarray}\label{eq:finish}
\E\|f(x)-f(y)\|_p^p&\ge& \sum_{i\in I} \sum_{t\in \mathbb
Z\cap\nonumber
[s_{i-1},s_i]}\E |d(x,W^i_t)-d(y,W^i_t)|^p\\
&\ge& [\Omega(\delta_{\varepsilon}(x,y) \cdot d(x,y))]^p
\sum_{i=-6}^3\nonumber
|Z\cap [s_{i-1},s_i]|\\
&\ge&[\Omega(\delta_{\varepsilon}(x,y) \cdot
d(x,y))]^p\cdot\frac{s_3-s_{-7}}{2}\\
&\ge&[\Omega(\delta_{\varepsilon}(x,y) \cdot d(x,y))]^p \cdot
V_\mu(x,y)\nonumber ,
\end{eqnarray}
where in \eqref{eq:finish} we used the fact that \eqref{eq:noloss}
implies that $s_3-s_{-7}\ge 1$.

This completes the proof of Lemma~\ref{lem:main}. \qed

\smallskip

\begin{remark}\label{rem:one-sided}
The above proof actually yields an embedding $\varphi$,
such that for all $x,y\in X$ satisfying \eqref{eq:noloss},
$$
\left[\log\frac{\mu (B(x,2d(x,y)))}{\mu
(B(x,d(x,y)/512))}\right]^{1/p}\cdot \delta_{\varepsilon}(x, y)
 \leq \frac{||\varphi(x) - \varphi(y)||_p}{d(x,y)} \leq C \left[\log
 \Phi(\mu)\right]^{1/p}.
$$
\end{remark}

\smallskip

\begin{remark}\label{remark:chernoff}
If in the above proof we use sampling and a standard
Chernoff bound instead of taking expectations, we can ensure that
the embedding takes values in $\mathbb R^k$, where $k=O[(\log
n)\log\Phi(\mu)]$.
(This is because the lower bound on $\E|d(x,W^i_t)-d(y,W^i_t)|^p$ relies
on an event that happens with constant probability, similar to \cite{LLR95}.)
In particular, when $\mu$ is the counting
measure on $X$ we get that $k=O[(\log n)^2]$. It would be
interesting to improve this bound to $k=O(\log n)$ (if $p=2$ then
this follows from the Johnson-Lindenstrauss dimension reduction
lemma~\cite{jl}).
\end{remark}

\subsection{Optimal volume-respecting
embeddings}\label{section:volume}

Here we prove Theorem~\ref{thm:feige}.  Let $f:X\to \ell_2$ be the
embedding constructed in the previous section and
$g=f/\sqrt{50\log \Phi(\mu)}$, so that $g$ is $1$-Lipschitz. For
concreteness, we denote by $(\Omega,\Pr)$ the probability space over
which the random embedding $g$ is defined.

\begin{lemma}\label{lem:distance}
Fix a subset $Y\subseteq X$, $x \in X\setminus Y$ and let $y_0\in Y$
satisfy $d(x,Y)=d(x,y_0)$. Let $Z$ be any $\ell_2$ valued random
variable on $(\Omega,\Pr)$ which is measurable with respect to the
$\sigma$- algebra generated by the random variables $\{g(y)\}_{y\in
Y}$. Then
$$
\frac{\sqrt{\E ||Z - g(x)||_2^2}}{d(x,Y)} \geq
C\cdot\delta_\e(x,y_0)\cdot\sqrt{ \frac{1}{\log
\Phi(\mu)}\left\lfloor\log\left(\frac{\mu(B(x,2d(x,y_0)))}{\mu(B(x,d(x,y_0)/512))}\right)\right\rfloor},
$$
where $C$ is a universal constant.
\end{lemma}
\begin{proof}
 We use the notation of
Section~\ref{section:proofbourgain}. Denote $R=\frac14 d(x,y_0)$ and
write $Z=(Z_t^i:\ i\in I, \ t\in J)$. Consider the events
$\widetilde{\mathcal{E}}_{i,t}^{\mathrm{big}}=\left\{Z_t^i\ge
\frac{1}{512} \delta_{\varepsilon}(x,y_0) R\right\}$ and
$\widetilde{\mathcal{E}}_{i,t}^{\mathrm{small}}= \left\{Z_t^i<
\frac{1}{512} \delta_{\varepsilon}(x,y_0) R\right\}$. Arguing as in
Section~\ref{section:proofbourgain}, it is enough to check that
$\mathcal E_3$ is independent of
$\widetilde{\mathcal{E}}_{i,t}^{\mathrm{big}}$ and that $\mathcal
E_2$ is independent of $\mathcal E_1\cap
\widetilde{\mathcal{E}}_{i,t}^{\mathrm{small}}$. Observe that the
proof of assertions \ref{item:firstind} and \ref{item:secondind} in
Claim~\ref{claim:probabilistic} uses only the fact that $d(x,y)\ge
4R$ (when considering $z\in B(y,\frac{1}{512}
\delta_{\varepsilon}(x,y) R)$), and this now holds for all $y\in Y$.
Since we assume that $Z^i_t$ is measurable with respect to
$\{d(y,W_t^i)\}_{y\in Y}$, the required independence follows.
\end{proof}

We remark that we will apply Lemma~\ref{lem:distance} to random
variables of the form $Z=\sum_{y\in Y} c_y g(y)$, where the $c_y$'s
are scalars. However, the same statement holds for $Z$'s which are
arbitrary functions of the variables $\{g(y)\}_{y\in Y}$.

The following lemma is based on a Bourgain-style embedding.

\begin{lemma}\label{lem:small}
 Define a
subset $S\subseteq X$ by $S= \{ x \in X : |B(x,R)| \leq e|B(x, R/2)|
\}$. Then there exists a 1-Lipschitz map $F : X \to L_2$ such that
if $x \in S$ and $Y \subseteq X$ with $d(x,Y) \geq R$, then
$$
d_{L_2}\left(F(x), \mathrm{span} \{ F(y) \}_{y \in Y}\right) \ge
\frac{C' R}{\sqrt{\log n}},
$$
where $C' > 0$ is a universal constant.
\end{lemma}

\begin{proof}
For each $t \in \{1,2,\ldots,\lceil \log n \rceil \}$, let $W_t
\subseteq X$ be a random subset which contains each point of $X$
independently with probability $e^{-t}$. Let $g_t(x) =
\min\{d(x,W_t),R/4\}$ and define the {\em random} map $f =
\frac{1}{\sqrt{\lceil \log n\rceil}} \left(g_1 \oplus \cdots \oplus
g_{\lceil \log n \rceil}\right)$ so that $||f||_\Lip \leq 1$.
Finally, we define $F : X \to L_2(\nu)$ by $F(x) = f(x)$, where
$\nu$ is the distribution over which the random subsets $\{W_t\}$
are defined.

Now fix $x \in S$ and let $t \in \mathbb N$ be such that $e^t \leq
|B(x,R/2)| \leq e^{t+1}$. Let $\mathcal E_{\mathrm{far}}$ be the
event $\left\{d(x,W_t) \geq R/4\right\}$ and let $\mathcal
E_{\mathrm{close}}$ be the event $\left\{d(x, W_t) \leq
R/8\right\}$. Clearly both such events are independent of the values
$\{ g_t(y) : d(x,y) \geq R \} \supseteq \{g_t(y)\}_{y \in Y}$ (this
relies crucially on the use of $\min \{ \cdot, R/4 \}$ in the
definition of $g_t$). Fixing $c_y \in \mathbb R$ for each $y \in Y$,
we see that
\begin{eqnarray*}
\left\| F(x) - \sum_{y \in Y} c_y F(y)\right\|^2_{L_2(\nu)}
&=& \mathbb E_\nu\left(f(x) - \sum_{y \in Y} c_yf(y)\right)^2 \\
&\ge&\frac{1}{2\log n} \mathbb E_\nu \left(g_t(x) - \sum_{y \in Y} c_yg_t(y)\right)^2 \\
&=& \Omega\left(\frac{R^2}{\log n} \cdot {\min \big\{ \nu(\mathcal
E_{\mathrm{far}}), \nu(\mathcal E_{\mathrm{close}}) \big\}}\right).
\end{eqnarray*}
Finally, we observe that by the definition of $S$, $\nu(\mathcal
E_{\mathrm{far}})$ and $\nu(\mathcal E_{\mathrm{close}})$ can
clearly be lower bounded by some universal constant.
\end{proof}

\medskip

\begin{proof}[Proof of Theorem~\ref{thm:feige}]
Using the notation of Lemma~\ref{lem:distance}, consider the Hilbert
space $H=L_2(\Omega,\ell_2)$, i.e. the space of all square
integrable $\ell_2$ valued random variables $\zeta$ on $\Omega$
equipped with the Hilbertian norm
$\|\zeta\|_2=\sqrt{\E\|\zeta\|_2^2}$. Defining $G:X\to H$ via
$G(x)=g(x)$, Lemma~\ref{lem:distance}  implies that for every
$Y\subseteq X$ and $x\in X\setminus Y$,
\begin{eqnarray}\label{eq:L_2(Omega)}
\frac{d_H(G(x),\mathrm{span}(\{G(y)\}_{y\in Y}))}{d(x,Y)}\ge
C\cdot\delta_\e(x,y_0)\cdot\sqrt{ \frac{1}{\log
\Phi(\mu)}\left\lfloor\log\left(\frac{\mu(B(x,2d(x,y_0)))}{\mu(B(x,d(x,y_0)/512))}\right)\right\rfloor}.
\end{eqnarray}

We now argue as in the proof of Theorem~\ref{thm:pad}. Let $H_1,
H_2$ be Hilbert spaces and $G_1: X\to H_1$, $G_2:X\to H_2$ be two
$1$-Lipschitz mappings satisfying for every $Y\subseteq X$ and
$x\in X\setminus Y$,
\begin{multline*}
\frac{d_{H_1}(G_1(x),\mathrm{span}(\{G_1(y)\}_{y\in
Y}))}{d(x,Y)}\\\ge
\frac{C}{16+16\log\left(\frac{|B(x,2d(x,y_0))|}{|B(x,d(x,y_0)/512)|}\right)}
\sqrt{ \frac{1}{\log
n}\left\lfloor\log\left(\frac{|B(x,2d(x,y_0))|}{|B(x,d(x,y_0)/512)|}\right)\right\rfloor},
\end{multline*}
and
$$
\frac{d_{H_2}(G_2(x),\mathrm{span}(\{G_2(y)\}_{y\in Y}))}{d(x,Y)}\ge
\frac{C}{\alpha_X} \sqrt{ \frac{1}{\log
n}\left\lfloor\log\left(\frac{|B(x,2d(x,y_0))|}{|B(x,d(x,y_0)/512)|}\right)\right\rfloor},
$$
where $d(x,y_0)=d(x,Y)$, and we used~\eqref{eq:L_2(Omega)} with
$\mu$ being the counting measure on $X$. Also, by
Lemma~\ref{lem:small} there is a Hilbert space $H_3$ and a mapping
$G_3:X\to H_3$ such that for every such $Y,x,y_0$
$$
\frac{d_{H_3}(G_3(x),\mathrm{span}(\{G_3(y)\}_{y\in Y}))}{d(x,Y)}\ge
\frac{C'}{\sqrt{\log n}}\cdot {\bf 1}_{\{|B(x,d(x,y_0)/2)|\leq e
|B(x,d(x,y_0)/8)|\}}.
$$

Denoting $H=H_1\oplus H_2 \oplus H_3$ and $G=\frac{1}{\sqrt{3}}(G_1\oplus
G_2 \oplus G_3)$, similar argument as in the proof of Theorem~\ref{thm:pad}
implies that
$$
\frac{d_{H}(G(x),\mathrm{span}(\{G(y)\}_{y\in Y}))}{d(x,Y)}\ge
\Omega\left(\frac{1}{\sqrt{\alpha_X\log n}}\right).
$$
Now, Feige's argument (see the proof of Lemma 15 in
\cite{Feige00}) yields the required result.
\end{proof}

\smallskip
\begin{remark}
Note that, by general dimension reduction techniques
which preserve distance to affine hulls \cite{Magen02}, the dimension
of the above embedding can be reduced to $O(k \log n$) while
maintaining the volume-respecting property for $k$-point subsets.
\end{remark}

\section{Low-dimensional embeddings of planar metrics}\label{section:planar}

In this section we refine the ideas of the previous section and
prove Theorem~\ref{thm:planar}. We say that a metric $(X,d)$ is
planar (resp. excludes $K_{s,s}$ as a minor) if there exists a
graph $G=(X,E)$ with positive edge weights, such that $G$ is
planar (resp. does not admit the complete bipartite graph
$K_{s,s}$ as a minor) and $d(\cdot,\cdot)$ is the shortest path
metric on a subset of $G$. We shall obtain optimal low-dimensional
embeddings of planar metrics into $\ell_\infty$ by proving the
following more general result.

\begin{theorem}
\label{thm:yuri} Let $(X,d)$ be an $n$-point metric space that excludes
$K_{s,s}$ as a minor. Then $X$ embeds into
$\ell_\infty^{O(3^s(\log s) \log n)}$ with distortion $O(s^2)$.
\end{theorem}

We will need three lemmas.
The first one exhibits a family of decompositions with respect to a
diameter bound $\Delta>0$; it follows easily from \cite{KPR93},
with improved constants due to \cite{FT03}.
Note that in contrast to Definition~\ref{def:padded}
(and also to Rao's embedding \cite{Rao99}),
we require that $x$ and $y$ are padded simultaneously.

\begin{lemma}
\label{lem:kpr} There exists a constant $c$ such that
for every metric space $(X,d)$ that excludes $K_{s,s}$ as a minor,
and for every $\Delta > 0$, there exists a set of $k = 3^s$
partitions $P_1, \ldots, P_k$ of $X$, such that
\begin{enumerate}
\setlength{\itemsep}{0in}
\item For every $C \in P_i$, $\diam(C) < \Delta$. \item For
every pair $x,y \in X$, there exists an $i$ such that for
$T=cs^2$,
$$B(x,\Delta/T) \subseteq P_i(x) \textrm{ {\bf and }}
B(y,\Delta/T) \subseteq P_i(y).$$
\end{enumerate}
\end{lemma}

\begin{proof}
Fix a edge-weighted graph $G$ that does not admit $K_{s,s}$ as a
minor and whose shortest path distance is $d(\cdot,\cdot)$. Fix
also some $x_0\in X$ and $\delta>0$. For $i\in \{0,1,2\}$ and
$j\in \{0\}\cup\mathbb N$ define:
$$
A^i_j=\left\{x\in X:\ 9(j-1)+3i\le \frac{d(x,x_0)}{\delta} < 9j+3i\right\}.
$$
For every $i$, $P^i=\{A^i_j\}_{j\ge 0}$ clearly forms a partition of $X$.
Let us say that a subset $S\subseteq X$ cuts a subset
$S'\subseteq X$ if $S\cap S'\neq \emptyset$ and $S'\not \subseteq
S$.
Observe that for every $x\in X$ at most one of the sets
$\{A^i_j:\ i=0,1,2\ ;j=0,1,2,\ldots\}$ cuts $B(x,\delta)$,
as otherwise there exist $z_1,z_2\in B(x,\delta)$ for which
$d(z_1,z_2) \ge d(x_0,z_1)-d(x_0,z_2) \ge 3\delta$.
Thus,
for every $x,y\in X$, for one of the partitions $P^0,P^1,P^2$ both
$B(x,\delta)$ and $B(y,\delta)$ are contained in one of its
clusters. For each cluster $C$ of the partitions $P^0,P^1,P^2$,
consider the subgraph of $G$ induced on the points of $C$,
partition $C$ into its connected components,
and apply the above process again to each such connected component.
Continuing this way a total of $s$ times, we end up with $3^s$ partitions,
and in at least one of them, neither $B(x,\delta)$ nor $B(y,\delta)$ is cut.
The results of \cite{KPR93, FT03} show there exists a constant
$c>0$ such that the diameter of each cluster in the resulting partitions
is at most $cs^2 \delta$,
and the lemma follows by setting $\delta=\Delta/(cs^2)$.
\end{proof}

We next consider a collection of such decompositions, with diameter
bounds $\Delta>0$ that are proportional to the integral powers of $4T$.
Furthermore, we need these decompositions to be nested.

\begin{lemma}\label{lem:nested}
Let $(X,d)$ be a metric space that excludes
$K_{s,s}$ as a minor, and let $T=O(s^2)$ be as in Lemma~\ref{lem:kpr}. Then for
every $a>0$ there exists $k=3^s$ families of partitions of $X$,
$\{P_u^i\}_{u\in \mathbb Z}$, $i=1,\ldots, k$ with the following
properties:
\begin{enumerate}
\setlength{\itemsep}{0in}
\item For each $i$ the partitions $\{P_u^i\}_{u\in \mathbb Z}$ are
nested, i.e. $P^i_{u-1}$ is a refinement of $P^i_u$ for all $u$.
\item For each $i$, every $C\in P_u^i$ satisfies $\diam(C) < a(4T)^u$.
 \item For
each $u\in \mathbb Z$ and every pair $x,y \in X$, there exists an
$i$ such that,
$$B(x,a (4T)^{u}/(2T)) \subseteq P_u^i(x) \textrm{ {\bf and }}
B(y,a (4T)^{u}/(2T)) \subseteq P_u^i(y).$$
\end{enumerate}
\end{lemma}

\begin{proof} Let $P_1,\ldots,P_k$ be partitions as in
Lemma~\ref{lem:kpr} with $\Delta=a(4T)^u$ and let $Q_1,\ldots,Q_k$ be
partitions as in Lemma~\ref{lem:kpr} with $\Delta=a(4T)^{u-1}$.
Fix $j$ and $C\in P_j$, let $S_C=\{A\in Q_j:\ A\cap C\neq
\emptyset,\ \mathrm{but}\ A\not\subseteq C\}$, and replace every
$C \in P_j$ by the sets $A \in S_C$ and the set $ C' = C \setminus
\bigcup_{A \in S_C} A$. Continuing this process we replace the
partition $P_j$ by a new partition $P_j'$ such that $Q_j$ is a
refinement of $P_j'$. Note that we do not alter $Q_j$.
Since $\diam(A) \leq a(4T)^{u-1}$, we have that
if $C \in P_j$ and $B(x, a(4T)^{u}/T) \subseteq C$,
then $B(x, 2a(4T)^{u-1}) \subseteq C'$. Continuing
this process inductively we obtain the required families of nested
partitions.
\end{proof}

We next use a nested sequence of partitions $\{P_u\}_{u\in \mathbb Z}$
to form a mapping $\psi:X \to {\mathbb R}^{O(\log |X|)}$.
\begin{lemma} \label{lem:onemap}
Let $\{P_u\}_{u\in \mathbb Z}$ be a sequence of partitions of $X$
that is nested (i.e. $P_{u-1}$ is a refinement of $P_u$),
and let $m\ge 0$ and $D\ge 2$ be such that for all $C\in P_u$,
$\diam (C) < 2^mD^u$. Assume further that $P_{u_1}=\{X\}$,
$P_{u_2}=\{\{x\}:\ x\in X\}$. Then for all $u_2\le u\le u_1$ and all
$A\in P_u$ there exists a mapping
$\psi: A\to \mathbb R^{2\lceil \log_2 |A|\rceil}$ that satisfies:
\begin{enumerate}
\setlength{\itemsep}{0in}
\renewcommand{\theenumi}{\rm{(\alph{enumi})}}
\item For every $x\in A$ and every $1\le j\le 2\lceil \log_2
|A|\rceil$ there exists $u'<u$ for which $|\psi(x)_j|=
\min\{d(x,X\setminus P_{u'}(x)),2^m D^{u'}\}$,
\item For all $x,y\in A$, $||\psi(x) -
\psi(y)||_{\infty} \leq 2\,d(x,y)$,
\item  If $x,y \in A$ are such
that for some $u'\le u$, $d(x,y)\in [2^mD^{u'-1},2^{m+1}D^{u'-1})$
and there exists a cluster $C\in P_{u'}$ for which $x,y\in C$,
$B(x, 2^{m+1}D^{u'-2}) \subseteq P_{u'-1}(x)$ and $B(y,
2^{m+1}D^{u'-2}) \subseteq P_{u'-1}(y)$, then $||\psi(x) -
\psi(y)||_{\infty} \geq \frac{d(x,y)}{2D}$.
\end{enumerate}
\end{lemma}

\begin{proof}
Proceed by induction on $u$. The statement is vacuous for $u=u_2$,
so we assume it holds for $u$ and construct the
required mapping for $u+1$. Fix $A\in P_{u+1}$ and assume that
$H=\{A_1,\ldots, A_r\}\subseteq P_u$ is a partition of $A$. By
induction there are mappings $\psi_i: A_i\to \mathbb R^{2\lceil
\log_2 |A_i|\rceil}$ satisfying (a)-(c) above (with respect to $A_i$ and $u$).
\remove{
\begin{enumerate}
\setlength{\itemsep}{0in}
\item For every $x\in A_i$ and every $1\le j\le 2\lceil \log_2
|A_i|\rceil$ there exists $u'<u$ for which $|\psi_i(x)_j|=
d(x,X\setminus P_{u'}(x))$, \item For all $x,y\in A_i$,
$||\psi_i(x) - \psi_i(y)||_{\infty} \leq 2\,d(x,y)$, \item  If
$x,y \in A_i$ are such that for some $u'\le u$ there exists a
cluster $C\in P_{u'}$ for which $x,y\in C$, $d(x,y) \in
[2^mD^{u'-1}, 2^{m+1}D^{u'-1})$, $B(x, 2^{m+1}D^{u'-2}) \subseteq
P_{u'-1}(x)$ and $B(y, 2^{m+1}D^{u'-2}) \subseteq P_{u'-1}(y)$,
then $||\psi_i(x) - \psi_i(y)||_{\infty} \geq \frac{d(x,y)}{2D}$.
\end{enumerate}
} 

For $h \in \mathbb N$ denote $\mathcal C_h = \{ A_i \in H :
2^{h-1} < |A_i| \leq 2^{h}\}$. We claim that for every
$i=1,\ldots,r$ there is a choice of a string of signs $\sigma^i\in
\{-1,1\}^{2\lceil \log_2 |A|\rceil-2\lceil \log_2 |A_i|\rceil}$
such that for all $h$ and for all distinct $A_i,A_j\in \mathcal
C_h$, $\sigma^i\neq \sigma^j$. Indeed, fix $h$; if $h\ge \log_2
|A|$ then for $A_i\in \mathcal C_h$, $|A_i|>2^{h-1} \ge |A|/2$;
thus $|\mathcal C_h|=1$ and there is nothing to prove. So, assume
that $h<\log_2 |A|$ and note that $|\mathcal C_h|\le |A|/2^{h-1}$.
Hence, the required strings of signs exist provided $2^{2\lceil
\log_2 |A|\rceil-2h}\ge |A|/2^{h-1}$, which is true since
$h+1\le \lceil \log_2 |A|\rceil$.

Now, for every $i=1,\ldots,r$ define a mapping $\zeta_i:
A_i\to \mathbb R^{2\lceil \log_2 |A|\rceil-2\lceil \log_2 |A_i|\rceil}$ by
$$\zeta_i(x)=\min\{d(x,X\setminus A_i),2^mD^u\}\cdot\sigma^i.$$
Finally, define
the mapping $\psi: X\to \mathbb R^{2\lceil \log_2 |A|\rceil}$ by
$\psi|_{A_i}=\psi_i\oplus \zeta_i$. Requirement (a) holds for
$\psi$ by construction. To prove requirement (b), i.e. that $\psi$ is
$2$-Lipschitz, fix $x,y\in A$. If for some $i$, both $x,y\in A_i$ then
by the inductive hypothesis $\psi_i$ is $2$-Lipschitz, and clearly
$\zeta_i$ is $1$-Lipschitz, so $\|\psi(x)-\psi(y)\|_\infty \le 2 d(x,y)$.
Otherwise, fix a coordinate $1\le j\le 2\lceil \log_2
|A|\rceil$ and use (a) to take $u'\le u$ such that $|\psi(x)_j|=d(x,
X\setminus P_{u'}(x))$; since $y\notin P_{u'}(x)$, this is at most $d(x,y)$.
It similarly follows that $|\psi(y)_j| \le d(x,y)$, and hence
$|\psi(x)_j-\psi(y)_j|\le 2d(x,y)$.

To prove that requirement (c) holds for $\psi$, take $x,y\in A$ and
$u'\le u+1$ such that $d(x,y)\in [2^mD^{u'-1},2^{m+1}D^{u'-1})$
and there exists a cluster $C\in P_{u'}$ for which $x,y\in C$,
$B(x, 2^{m+1}D^{u'-2}) \subseteq P_{u'-1}(x)$ and $B(y,
2^{m+1}D^{u'-2}) \subseteq P_{u'-1}(y)$. The case $u'\le u$ follows by
induction, so assume that $u'=u+1$. Let $i,j\in \{1,\ldots,r\}$ be such that
$x\in A_i$, $y\in A_j$; then $i\neq j$,
since $\diam(A_i)<2^mD^u\le d(x,y)$.
Assume first $\lceil \log_2 |A_i|\rceil \ne\lceil \log_2 |A_j|\rceil$,
and without loss of generality suppose
$\lceil \log_2 |A_i|\rceil <\lceil \log_2 |A_j|\rceil$; then there
is a coordinate $\ell=2\lceil \log_2 |A_i|\rceil+1$ for which
\begin{eqnarray*}
|\psi(x)_\ell|=|\zeta_i(x)_1| = \min\{d(x,X\setminus A_i),2^m D^u\},
\end{eqnarray*}
and, for some $u'' < u$,
$$
|\psi(y)_\ell|=|\psi_j(y)_\ell| = \min\{d(y,X\setminus P_{u''}(y)),2^m D^{u''}\}.$$
It follows that $|\psi(x)_\ell|\ge 2^{m+1}D^{u-1}$
(since we assumed $B(x, 2^{m+1}D^{u-1}) \subseteq A_i$),
and that $|\psi(y)_\ell|\le 2^mD^{u-1}$, and therefore
$$|\psi(x)_\ell-\psi(y)_\ell|\ge 2^mD^{u-1}\ge \frac{d(x,y)}{2D}.$$
It remains to deal with the case $\lceil \log_2 |A_i|\rceil =\lceil
\log_2 |A_j|\rceil$. By our choice of sign sequences, in this case there is
an index $\ell$ for which $\sigma^i_\ell\neq \sigma^j_\ell$, and thus,
for $\ell'=\ell+2\roundup{\log_2|A_i|}$,
$|\psi(x)_{\ell'}-\psi(y)_{\ell'}| = |\psi(x)_{\ell'}| + |\psi(y)_{\ell'}|$.
Since we assumed $B(x, 2^{m+1}D^{u-1}) \subseteq A_i$
and $B(y, 2^{m+1}D^{u-1}) \subseteq A_j$, we get
$$|\psi(x)_{\ell'}-\psi(y)_{\ell'}| \ge 2^{m+2}D^{u-1}\ge \frac{2d(x,y)}{D}.$$
\end{proof}

Finally, we prove the main result of this section by a concatenating
several of the above maps $\psi$.

\begin{proof}[Proof of Theorem \ref{thm:yuri}]
For each $m\in \{0,1,\ldots, \lceil \log_2 (4cs^2)\rceil\}$
set $a=2^m$, apply Lemma~\ref{lem:nested} to obtain
$3^s$ families of nested partitions $\{P^m_{1,u}\}_{u\in \mathbb Z},\ldots
,\{P^m_{3^s,u}\}_{u\in \mathbb Z}$ that satisfy
the conclusion of Lemma~\ref{lem:nested} with $T=cs^2$.
For every $i=1,\ldots, 3^s$, let $\psi^m_i$ be the mapping
that Lemma~\ref{lem:onemap} yields for $\{P^m_{i,u}\}_{u\in\mathbb Z}$
when setting $A=X$ and $D=4cs^2$. Consider the map
$\Psi=\oplus_{m,i}\psi^m_i$, which takes values in
$\ell_\infty^{O(3^s(\log s)\log n)}$. Clearly $\Psi$ is
$2$-Lipschitz. Moreover, for every $x,y\in X$ there is $m\in
\{0,1,\ldots, \lceil \log_2 (4cs^2)\rceil\}$ and $u\in \mathbb Z$
such that $d(x,y) \in [2^mD^{u}, 2^{m+1}D^{u})$. By
Lemma~\ref{lem:nested}, there is $i\in \{1,\ldots, 3^s\}$ for which
$B(x, 2^{m+1}D^{u-1}) \subseteq P_{u}(x)$ and $B(y,
2^{m+1}D^{u-1}) \subseteq P_{u}(y)$; it then follows using
Lemma~\ref{lem:onemap} that
$$\|\Psi(x)-\Psi(y)\|_\infty \ge \|\psi^m_i(x)-\psi^m_i(y)\|_\infty =\Omega(d(x,y)/s^2),$$ as required.
\end{proof}

\subsection*{Acknowledgments}
The third author is grateful to Y. Bartal for discussions on
related issues during a preliminary stage of this work.


\bibliographystyle{abbrv}

\def\cprime{$'$}

\end{document}